\documentclass{aa}  

\usepackage{graphicx}
\usepackage{txfonts}
\usepackage{lipsum}
\usepackage{subcaption}         
                                
\usepackage{lscape}             
                                
\usepackage{placeins}

\usepackage{url}
\usepackage{hyperref}
\usepackage{orcidlink}
\usepackage{booktabs, geometry, multicol}

\newcommand {\bc}{\begin {center}}
\newcommand {\ec}{\end {center}}
\newcommand {\be}{\begin {equation}}
\newcommand {\ee}{\end {equation}}
\newcommand {\beq}{\begin {eqnarray}}
\newcommand {\eeq}{\end {eqnarray}}

\def\flux{erg\,s$^{-1}$\,cm$^{-2}$\xspace}
\def\lum{erg\,s$^{-1}$\xspace}

\def\nustar{{\it NuSTAR}\xspace}

\def\bepposax{{\it BeppoSAX}\xspace}
\def\xmm{{\it XMM-Newton}\xspace}
\def\chandra{{\it Chandra}\xspace}
\def\integral{{INTEGRAL}\xspace}
\def\asca{{ASCA}\xspace}
\def\rosat{{ROSAT}\xspace}
\def\swift{{Swift}\xspace}
\def\gaia{{Gaia}\xspace}

\newcommand{\fullsrc}{2RXP\,J130159.6$-$635806\xspace}
\newcommand{\src}{J1301\xspace}

\begin{document}
\title{Discovery of a bimodal luminosity distribution in  persistent Be/X-ray pulsar 2RXP\,J130159.6$-$635806}
\titlerunning{Bimodal accretion in 2RXP\,J130159.6$-$635806}

   \author{
   Alexander Salganik \inst{\ref{in:UTU}}\orcidlink{0000-0003-2609-8838}
   \and Sergey S. Tsygankov \inst{\ref{in:UTU}}\orcidlink{0000-0002-9679-0793} 
   \and Maria Chernyakova
   \inst{\ref{in:DCU}, \ref{in:DCIA}}\orcidlink{0000-0002-9735-3608}
   \and Denys Malyshev  \inst{\ref{in:UT}}\orcidlink{0000-0001-9689-2194}
   \and Juri Poutanen \inst{\ref{in:UTU}}\orcidlink{0000-0002-0983-0049}}

\authorrunning{Salganik et al.}
\institute{Department of Physics and Astronomy, FI-20014 University of Turku,  Finland \label{in:UTU} \\ \email{alsalganik@gmail.com}
\and
School of Physical Sciences and Centre for Astrophysics \& Relativity, Dublin City University, Glasnevin, D09 W6Y4, Ireland \label{in:DCU} 
\and
Dublin Institute for Advanced Studies, 31 Fitzwilliam Place, Dublin 2 \label{in:DCIA}
\and 
Institut f\"ur Astronomie und Astrophysik T\"ubingen, Universit\"at T\"ubingen, Sand 1, D-72076 T\"ubingen, Germany
\label{in:UT}
}

   \date{Received XXX}

\abstract {We present a comprehensive analysis of \fullsrc, a persistent low-luminosity Be/X-ray pulsar, focusing on its transition to a spin equilibrium state and the discovery of a bimodal luminosity distribution revealing possibly a new accretion regime. Using data from \nustar, \swift, \xmm, and \chandra observatories, we investigate changes in the pulsar's timing and spectral properties. After more than 20 years of continuous spin-up, the pulsar's spin period stabilized, marking the onset of spin equilibrium. This transition was accompanied by the emergence of a previously unobserved accretion regime at $L_{\rm bol} = (2.0_{-1.0}^{+2.3})\times 10^{34}$\,\lum, an order of magnitude lower than its earlier quiescent state. After that, the source occasionally switched between these regimes, remaining in each state for extended periods, with the transition time from a luminosity of $10^{35}$\,\lum to $10^{34}$\,\lum taking less than 2.3~day. The analysis of the spectral data  collected during this new low-luminosity state revealed a two-hump  shape which is different from the cutoff power-law spectra observed at higher luminosities. The discovery of pulsations in this state, together with the hard spectral shape, demonstrates ongoing accretion. We estimate the magnetic field strength to be $\sim 10^{13}$~G based on indirect methods. Additionally, we report a hint of a previously undetected $\sim$90-day orbital period in the system.}

\keywords{accretion, accretion disks -- magnetic fields -- pulsars: individual: 2RXP J130159.6$-$635806 -- stars: neutron -- X-rays: binaries}

\maketitle

\section{Introduction}

Accreting X-ray pulsars (XRPs) in high-mass X-ray binaries (HMXBs) are systems where a strongly magnetized neutron star accretes matter from a massive companion, typically an O or B-type star. Among them, Be/X-ray binaries (BeXRPs) are the most common subclass, consisting of a neutron star orbiting a Be-type companion surrounded by a circumstellar decretion disk \citep[for reviews, see][]{Reig2011, MushtukovTsygankov2024}. These systems are typically known for their transient nature, with periodic Type I outbursts that occur near periastron and reach luminosities up to $\sim10^{37}$ erg s$^{-1}$, and occasional Type II outbursts, which are giant, irregular events exceeding $10^{37}$\,\lum. Additionally, there is the possibility of the XRP abrupt fading due to the cessation of accretion when transitioning to the so-called `propeller regime', where accretion is inhibited by a centrifugal barrier (i.e., when the magnetospheric radius exceeds the co-rotation radius; \citealt{Illarionov1975}). XRPs can remain detectable in X-rays even in the propeller state, potentially due to emission from the surface of the neutron star \citep[see, e.g.,][for the cases of 4U\,0115$+$63 and V\,0332$+$53]{Tsygankov2016}. However, a small number of BeXRPs do not follow this pattern and instead exhibit persistent, low-level accretion over extended periods. 

These persistent low-luminosity BeXRPs exhibit X-ray luminosities of 
$\sim$$10^{34}$--$10^{35}$\,\lum, significantly lower than typical outbursting BeXRPs \citep{Reig1999, Reig2011}. They are characterized by long pulse periods, typically ranging from hundreds to thousands of seconds, with notable examples including X\,Persei, RX\,J0440.9$+$4431, and RX\,J0146.9$+$5121 \citep{Reig1999, Reig2011}. Their long-term stability suggests a distinct accretion mechanism, differing from the periastron-driven outbursts observed in transient BeXRPs.  One possible explanation is a so-called `cold disk' accretion \citep{Tsygankov2017}, where the material continuously reaches the neutron star at low mass accretion rates, enabling sustained X-ray emission. Nevertheless, X\,Persei, the prototypical persistent low-luminosity BeXRP, exhibits long-term flux variations and alternating spin-up/spin-down episodes over years to decades, likely due to changes in the decretion disk \citep[see, e.g.,][]{Lutovinov2012}. \citet{Nakajima2019} reported a $\sim$7-year quasi-periodic X-ray modulation, which is $\sim$10 times longer than its 250~day orbital period, without spectral or hardness variations, suggesting intrinsic accretion rate changes.  Notably, RX\,J0440.9$+$4431, previously considered a persistent low-luminosity BeXRP \citep{Reig1999}, was observed during a transition into a transient state, displaying bright outbursts reminiscent of classical BeXRBs \citep{Ferrigno2013, Salganik2023}.
These observations indicate that persistent low-luminosity BeXRPs exhibit more diverse and dynamic behavior than previously thought, requiring further study to understand their long-term evolution.

This work focuses on a low-luminosity persistent BeXRP first detected in early 2004 during a Galactic plane scan by \integral\ \citep{Chernyakova2004}. Archival data analysis revealed that the source had previously been discovered by the \rosat\ observatory as \fullsrc\  \citep[hereafter \src;][]{Chernyakova2004}, also known as IGR~J13020$-$6359 in the \integral\ catalog \citep{Bird2006, Revnivtsev2006}. Using \xmm\ data, an XRP with the spin period $P_{\rm spin}$ of approximately 700~s was discovered in the system \citep{Chernyakova2005} with a Be star proposed as the companion. The optical counterpart was identified by \citet{Masetti2006} with a star of spectral class B0.5Ve \citep{Coleiro2013}. Recent photometric distance estimates based on \gaia data placed the source at $5.4^{+2.5}_{-1.5}$~kpc  \citep{BailerJones2021}. Long-term monitoring of the system using data from \xmm, \bepposax, \integral, and \asca has established that the source consistently maintained a flux of $(2$--$3) \times 10^{-11}$\,\flux\ in the 2--10~keV range during MJD~49349--53056 (1993--2004). This corresponds to a 2--10~keV luminosity of $\sim (5$--$20) \times 10^{34}$\,\lum, assuming the distance derived from \gaia\ observations. This classifies the system as a persistent low-luminosity BeXRP \citep{Chernyakova2005}.

The persistent nature of \src, coupled with its relative proximity to us, makes it an interesting target for long-term monitoring. Long-term timing of the source revealed a sustained spin-up with $\dot{P}= -6\times10^{-8}$\,s\,s$^{-1}$ during the period MJD~49378--50700, followed by an increase in the spin-up rate to $\dot{P} = -2 \times10^{-7}$\,s\,s$^{-1}$ during MJD~50700--53056  \citep{Chernyakova2005}. Interestingly, subsequent observations showed that this spin-up trend continued until MJD~56832 (2014) at the same rate \citep{Krivonos2015}. Thus, the XRP exhibited long-term spin-up for over 20 years.

The first detected outburst from \src was observed around MJD~53020 (2004) with a characteristic decay time of approximately 7.5~days, as seen by \integral\ \citep{Chernyakova2005}. During the outburst, the flux peaked at $\sim10^{-10}$\,\flux\ in the 2--10~keV range, while the spectrum demonstrated a typical cutoff power-law shape. Throughout all observations, both during and outside the outburst, the hydrogen column density remained nearly constant at $N_{\rm H}=(2.48\pm0.07)\times10^{22}~{\rm cm}^{-2}$ \citep{Chernyakova2005}.  

A decade later, a broadband spectral study of \src\ was conducted using \nustar\ (hereafter NuSTAR2014), confirming the cutoff power-law shape at a 2--10~keV flux of $3\times10^{-11}$\,\flux, consistent with the previously observed quiescent state \citep{Krivonos2015}. Additionally, an iron line at 6.4~keV was detected. The magnetic field of \src\ has not been estimated, and no cyclotron resonant scattering features (CRSFs) have been identified in its spectrum \citep{Krivonos2015}.

\begin{figure*}
\centering
\includegraphics[width=0.9\linewidth]{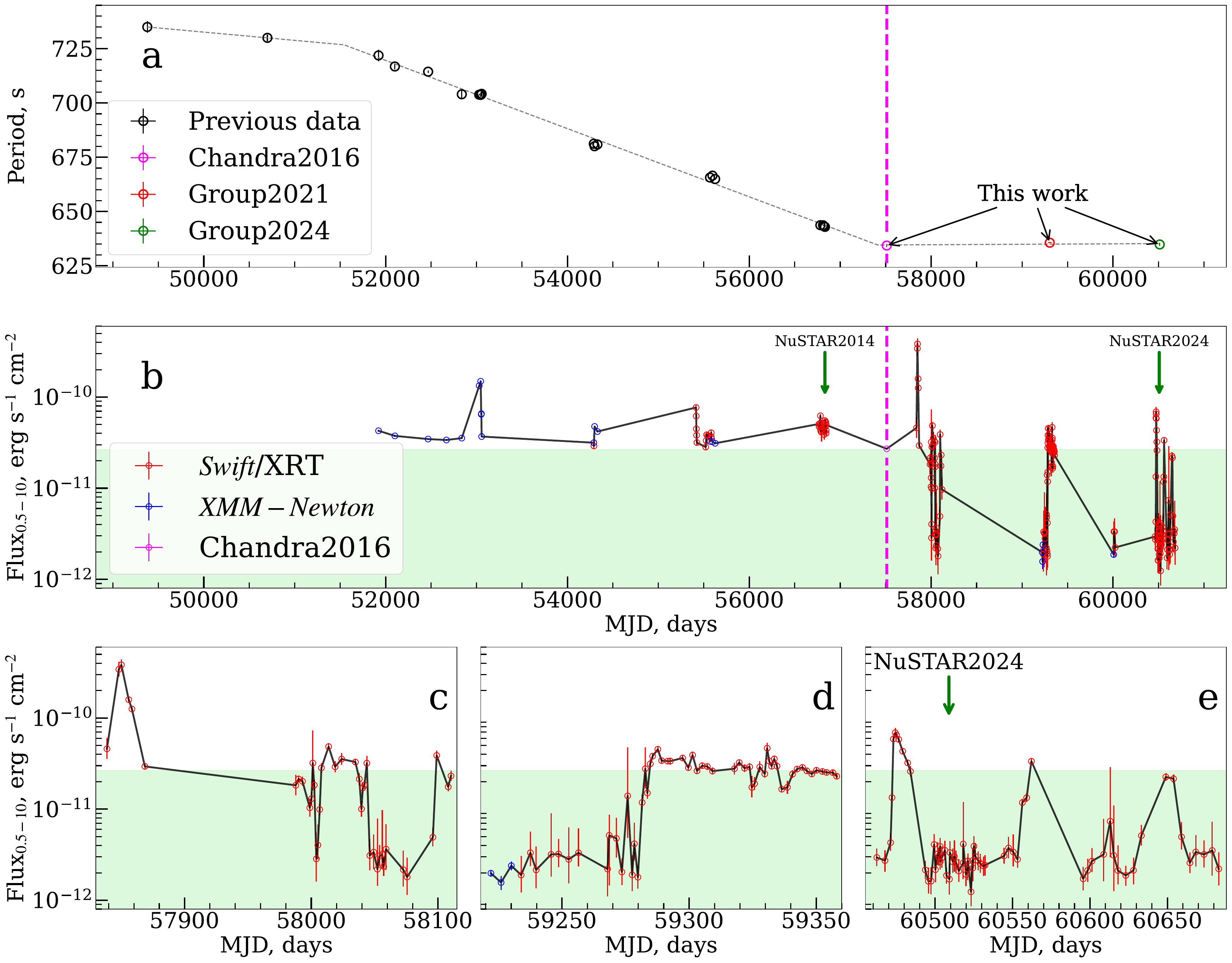}
\caption{\textit{Top}: Evolution of the pulse period over time (see Sect.~\ref{sec:timing} for details). The dotted line represents a three-piece broken curve approximation of the data points.  The dashed magenta line marks the transition to the equilibrium state. \textit{Middle}: Long-term 0.5--10~keV light curve of \src.  \textit{Bottom}: Zoomed-in segments of the long-term light curve after the transition. The light green zone indicates a flux range lower than the pre-transition quiescent flux. Green arrows indicate the times of the \nustar observations.}
\label{fig:equilibrium}
\end{figure*} 

In this study, we utilize data from the \nustar, \swift, \xmm, and \chandra X-ray observatories to examine the long-term timing and spectral characteristics of \src, focusing on its spin evolution. We report the discovery of the transition to an equilibrium state, accompanied by an emergence of new low-luminosity accretion regime.

\section{Data} 
\label{sec:data}
\subsection{\nustar}
\nustar consists of two identical coaxial X-ray focal plane modules (FPMs), FPMA and FPMB \citep{Harrison2013}, and is capable of reflecting X-rays up to 79 keV. 
\nustar has observed \src\ on 2024 July 17 with an effective exposure time of 30 ks (ObsID 31001007002, thereafter NuSTAR2024). The \nustar data were processed according to the official guidelines.\footnote{\url{https://heasarc.gsfc.nasa.gov/docs/nustar/analysis/nustar_swguide.pdf}} The \textsc{heasoft} package version 6.33.2 and calibration files \textsc{caldb} version 20240715 (clock correction file v186) were used for data processing. Spectra and light curves were extracted using the \texttt{nuproducts} procedure, part of the \texttt{nustardas} pipeline. The event files for both FPMA and FPMB were barycentrically corrected using the \texttt{barycorr} procedure. The data were extracted using a circular source region with a radius of 40\arcsec\ and a circular background region with a radius of 90\arcsec. The resulting light curves were co-added after background subtraction using the \texttt{lcmath} task in order to improve the statistics. No corrections were applied for binary motion, as the orbit of the source is unknown.

Additionally, we incorporated the observation of \src\ from \citet{Krivonos2015}, conducted on 2014 June 24 (ObsID 30001032002, thereafter NuSTAR2014), in our analysis for comparison with the broadband spectrum of NuSTAR2024. For data extraction, a source region with a radius of 60\arcsec\ and a background region with a radius of 90\arcsec\ were utilized.

\begin{table*}
\centering
\caption{Pulse period measurements obtained in this work.}
\begin{tabular}{lcccc} 
\hline\hline
ObsID & MJD range & Exposure & Flux$_{0.5-10}$  & Period  \\ 
 & (d) & (ks) &   ($10^{-11}$ erg s$^{-1}$ cm$^{-2}$) & (s)  \\ 
\hline
Chandra2016              & 57511.4--57511.6 & 9.8 &$2.71\pm0.05$&  $634.3 \pm 0.8$  \\ 
Group2021             & 59285.6--59309.2 & 38.8 & $3.55\pm0.95$ & $635.60 \pm 0.01$  \\ 
Group2024 & 60507.3--60520.0 & 40.0 & $0.27\pm0.10$& $634.8 \pm 0.7$\\ \hline
\end{tabular}
\label{table:period}
\end{table*}

\subsection{\xmm}

We took observations from \xmm Science Archive, the list can be found in Table~\ref{table:xmm_obs}.
The \xmm data were reduced using the \xmm SAS (Science Analysis System) \citep{Gabriel2004} version 21.0.0. The event files were processed using tasks \texttt{emproc} and \texttt{epproc} for MOS and PN detectors, respectively. The event files were filtered, and light curves and spectra were extracted with \texttt{evselect}. Barycenter corrections were applied using \texttt{barycen}, and response files (RMF, ARF) were generated with \texttt{rmfgen} and \texttt{arfgen}. Light curves were corrected using \texttt{epiclccorr}. Observations from different modules were combined using \texttt{epicspeccombine} if present, and summed light curves were produced with \texttt{lcmath}. 

\subsection{\swift}

To analyze the evolution of the flux (see Fig.~\ref{fig:equilibrium}) and spectrum as well as timing properties over the long-term light curve, we utilized archival data along with the results of a recent monitoring campaign with the XRT telescope \citep{Burrows2005} onboard the \textit{Neil Gehrels Swift} Observatory \citep{Gehrels2004}, covering MJD 54290--60683. The list of \swift observations analyzed in this work is presented in Table~\ref{table:swift_obs}.  Observations were performed in the PC mode, except 00030966037, 00030966040 which were performed in the WT mode. 
Although ObsID 00097781003 was excluded from the light curve construction due to its low statistics, it was still included in the construction of the combined \swift spectrum for a simultaneous fit with NuSTAR2024 data in Sect.~\ref{sec:spectrum}.
Data analysis software\footnote{\url{https://www.swift.ac.uk/user_objects/}} \citep{Evans2009} provided by the UK \swift Science Data Center was used to extract the source spectrum for each observation. 

The event data were processed to extract the light curves. First, the event files were barycenter corrected using \texttt{barycorr}, then light curve extraction was performed using \texttt{xselect}, where a 50\arcsec\ region filter was applied and a bin size of 5.0~s was set. The extracted light curves were then combined into two groups using \texttt{lcurve} in order to detect periodicities: Group2021 (ObsIDs 00095944010-22) and Group2024 (00097781001, NuSTAR2024, 00097781004-6) containing  the \nustar observation (see Table~\ref{table:period}).  For joint timing analysis (search for periodicities using $\chi^2$ distributions and pulse profile construction) with \textit{Swift} observations, the count rates for NuSTAR2024 were adjusted by applying a conversion factor of 0.3283, representing the ratio between \textit{NuSTAR} (3--79 keV) and \textit{Swift}/XRT (0.2--10 keV) count rates. This factor was obtained using WebPIMMS\footnote{\url{https://heasarc.gsfc.nasa.gov/cgi-bin/Tools/w3pimms/w3pimms.pl}} based on the spectral parameters for \src\ (see Table~\ref{table:spectrum}).

\subsection{\chandra}

During \chandra\ \citep{2000SPIE.4012....2W} observations  of \nustar's serendipitous sources near the Galactic plane, \src\ was covered by one of the \chandra\ pointings (ObsID 18087, hereafter Chandra2016) on 2016 May 3 with a 9.8 ks exposure \citep{Tomsick2018}. The data were analyzed with \textsc{CIAO} (Chandra Interactive Analysis of Observations) v.4.16, using calibration files from \textsc{CALDB} v.4.11.5. A barycentric correction was applied using \texttt{axbary} to correct the photon arrival times. Light curves were extracted with \texttt{dmextract} using 35\arcsec\ source and 90\arcsec\ background circular regions, with a time binning of 3.3~s. Spectral extraction was performed using the \texttt{specextract} procedure.

\subsection{Spectral data approximation}
\label{sec:spectral_data}
The resulting spectra of the source from all instruments were rebinned to have at least 1 count per energy bin using \texttt{grppha} utility and fitted using W-statistics\footnote{\url{https://heasarc.gsfc.nasa.gov/xanadu/xspec/manual/XSappendixStatistics.html}} \citep{Wachter1979} in the \textsc{xspec} v.12.14.0h package \citep{Arnaud1996}. All errors are given at 1$\sigma$ confidence level unless otherwise specified. All quoted fluxes in this paper are unabsorbed fluxes, unless explicitly stated otherwise. 
This applies to values reported in Figs.~\ref{fig:equilibrium} and \ref{fig:hist} as well as  Tables~\ref{table:period} and \ref{table:spectrum}, and throughout the text.

\section{Results} 
\subsection{Long-term light curve}
\label{sec:long_lc}

\begin{figure}
\centering
\includegraphics[width=1.0\linewidth]{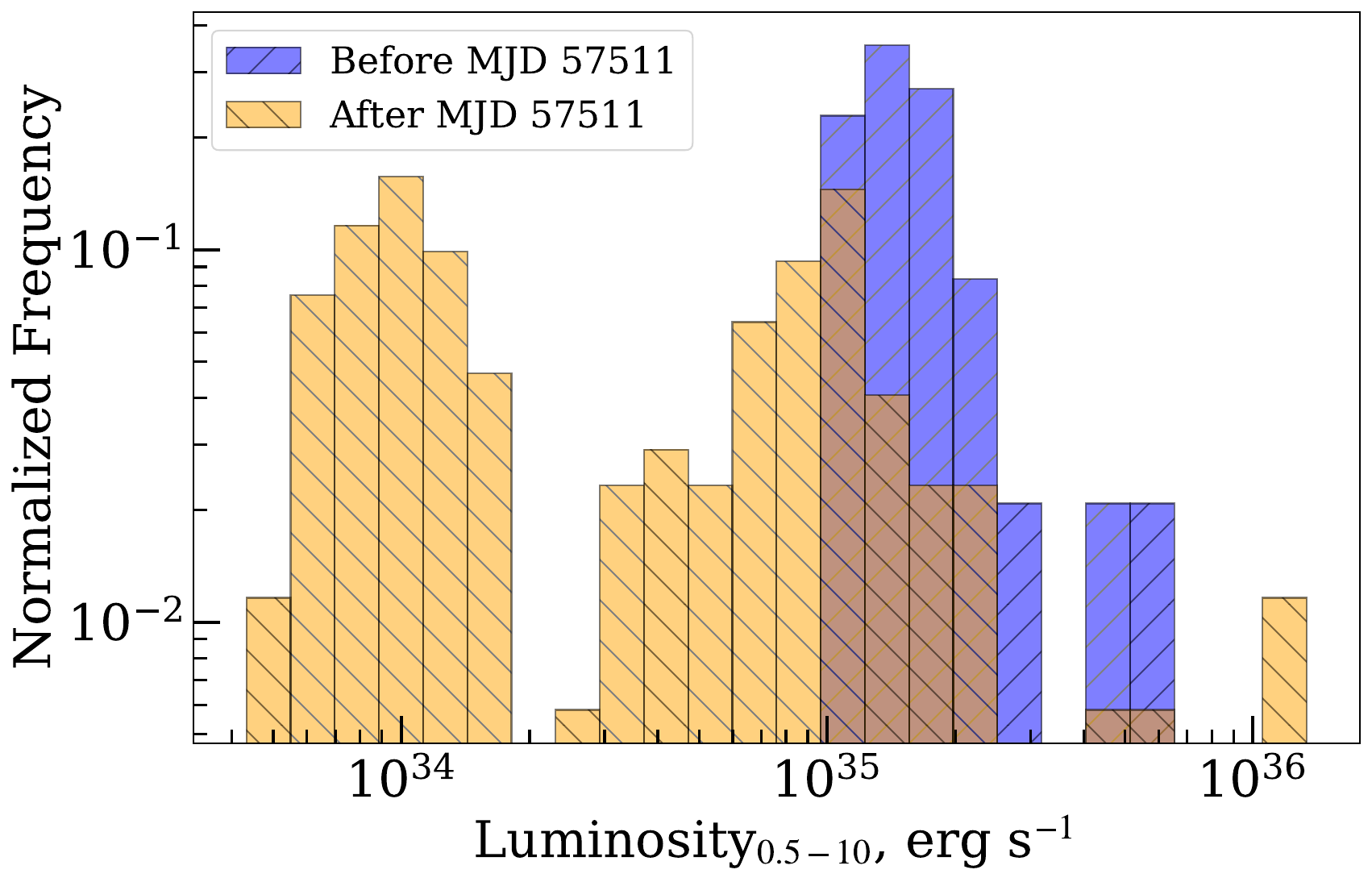}
\caption{Histogram of luminosity  distribution before and after the transition to equilibrium for \swift/XRT, \chandra, and \xmm observations. Fluxes were converted to luminosities assuming a distance of 5.4~kpc.}
 \label{fig:hist}
\end{figure} 

To construct the long-term light curve, \swift/XRT, \xmm and \chandra observations were fitted with a simple absorbed power-law model \texttt{tbabs}$\times$\texttt{po}, where the hydrogen column density $N_{\rm H}$ was frozen at $2.48 \times 10^{22}$~cm$^{-2}$ \citep{Chernyakova2005}. This choice was made due to the relatively low quality of individual spectra, which did not allow us to reliably constrain $N_{\rm H}$ for each observation separately. The long-term behavior of \src\ shows highly unusual activity for a BeXRP system.

Between MJD 51000 and 53000, \src\ maintained a stable flux, with the exception of a single outburst \citep{Chernyakova2005}. 
However, following the Chandra2016 observation (marked by the magenta dashed line in Fig.~\ref{fig:equilibrium}), \src\ began to display behavior atypical for BeXRPs: flux in the 0.5--10 keV range, which was normally around (2--4) $\times10^{-11}$\,\flux during the quiescent state, dropped to a few $\times10^{-12}$\,\flux on multiple occasions for extended periods. 
This indicated the emergence of a new stable state for \src, at a luminosity about ten times lower than the previously identified quiescent level (see the light-green stripe in Fig.~\ref{fig:equilibrium}). Hereafter, we refer to this state as the `lowest' state.

To investigate in more detail the change in variability before and after the Chandra2016 (MJD~57511) observation in greater detail, we constructed a histogram of the flux frequency, as shown in Fig.~\ref{fig:hist}. 
The histogram shows the distribution of the number of observations as a function of flux before and after MJD~57511. 
The data were divided into two intervals: observations taken before and after this date. 
Each data set was normalized separately by dividing the number of observations in each flux bin by the total number of observations across all flux bins in a given interval. 
This normalization allows for a clearer comparison between the two intervals. 

Before the transition, the fluxes clustered around luminosities of $\sim10^{35}$\,\lum in the 0.5--10 keV range. 
After the transition, a second cluster emerged at luminosities around $\sim10^{34}$\,\lum, indicating the presence of a new, lower stable state. 
Additionally, four significant outbursts were observed around 57850, 60470, 60560, 60650 as shown in Fig.~\ref{fig:equilibrium}.
Three consecutive outbursts around MJD 60470, 60560, and 60650 may suggest the presence of an orbital period, $P_{\rm orb}$, of about 90~days. 
However, the data quality does not allow for a high degree of confidence in this interpretation. 
If confirmed, this period would place \src\ among other Be/X-ray binaries on the \citet{Corbet1986} $P_{\rm spin}$--$P_{\rm orb}$ diagram.

To investigate whether the newly observed low-luminosity state is caused by increased absorption rather than intrinsic flux variations, we analyzed the broadband (0.5--79 keV) spectral properties of the source in both high ($\sim 10^{-11}$ \flux) and low ($\sim 10^{-12}$ \flux) flux states. This broader energy range provides better constraints on $N_{\rm H}$ compared to the 0.5--10 keV range. Specifically, we tested whether the hydrogen column density $N_{\rm H}$ might be significantly higher than the previously assumed value of $N_{\rm H} = 2.48 \times 10^{22}$ cm$^{-2}$ from \citet{Chernyakova2005}. The measured $N_{\rm H}$ in the low-luminosity state, $N_{\rm H} = (2.4 \pm 0.6) \times 10^{22}$ cm$^{-2}$ (using the \texttt{cutoffpl} model) or $N_{\rm H} = 1.6^{+1.5}_{-0.7} \times 10^{22}$ cm$^{-2}$ (using the \texttt{comptt+comptt} model), as listed in Table~\ref{table:spectrum}, remains consistent with the previously reported value within uncertainties. This confirms that the observed flux variations are not driven by changes in absorption but rather reflect intrinsic variations in the system's accretion rate.

Thus, after the Chandra2016 observation, the dynamic range of \src\ expanded significantly, spanning two orders of magnitude, with the new low state at the flux level of  $\sim10^{-12}$\,\flux and new outburst flux reaching up to $\sim10^{-10}$\,\flux.  
Crucially, rather than fading smoothly to a lower accretion level or displaying fast random variability, \src\ now switches between two well-defined accretion states, spending extended periods at both luminosities,  $\sim10^{35}$\,\lum and $\sim10^{34}$\,\lum.  
This clear bimodal behavior, in which the source oscillates between two stable luminosity states, suggests a change in its accretion regime after the Chandra2016 observation. Such bimodal behavior has so far never been observed in other BeXRPs, making \src currently a unique system (see Section~\ref{sec:low} for a further discussion).
\subsection{Timing analysis}
\label{sec:timing}

\begin{figure*}
\centering
\includegraphics[width=0.9\linewidth]{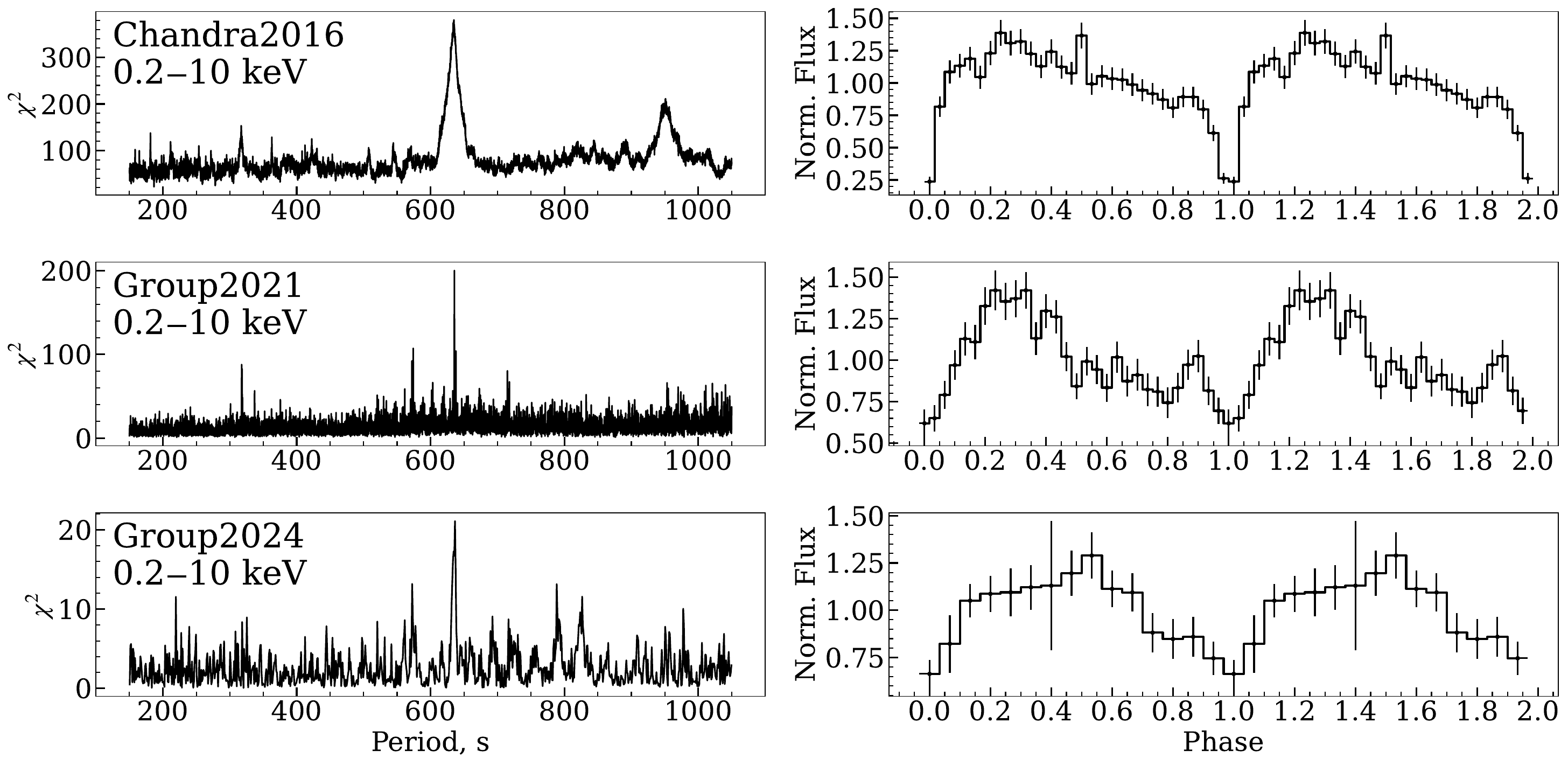}
\caption{\textit{Left}: 
$\chi^2$ periodograms for Chandra2016 (top panel), Group2021 (middle panel), and  Group2024 (bottom panel). 
\textit{Right}: Normalized pulse profiles of \src\ for corresponding intervals (see Sect.~\ref{sec:timing}).}
 \label{fig:period}
\end{figure*}

To obtain more details about the transition to a new accretion mode, we studied the rotational evolution of the XRP after this transition. 
For this purpose, we utilized the Chandra2016, Group2021, and Group2024 light curves, the latter being obtained when \src\ was in the `lowest' state.

To search for periodicities, we applied the epoch-folding technique, implemented in the \texttt{efsearch} utility from the \textsc{xronos} package. 
We successfully detected the presence of pulsations with high significance in all three observations (see Table~\ref{table:period} and Fig.~\ref{fig:period}). 
The uncertainty of the period was estimated by simulating a large number of light curves, where the source count rate $CR$ varied within the statistical error $ERR$: ${CR}_{\rm sim} = CR + {ERR} \times RAND([-1,1])$. 
For all simulated light curves, the pulse periods were measured to construct a distribution of their values. 
The mean of this distribution was taken as the final period for a given observation, and the standard deviation was used as uncertainty (see \citealt{Boldin2013}, for more details).

The spin period $P_{\rm spin}$ evolution of \src\ is presented in the upper panel of Fig.~\ref{fig:equilibrium}. 
As shown in the figure, after a long period of stable spin-up, it ceased around MJD~57500 (Chandra2016). 
After this, the spin period remained constant up to MJD~60500 (Group2024) with its derivative being $-8.5 \times10^{-8} < \dot{P} < 1.2\times10^{-8}$~s~s$^{-1}$. 
It is evident from the magenta line in Fig.~\ref{fig:equilibrium} that the abrupt change in the long-term light curve of \src coincided with the cessation of spin-up.

We also constructed pulse profiles for the Chandra2016, Group2021, and Group2024 observations. The Chandra2016 and Group2021 profiles demonstrate single-peaked shapes similar to those reported by \citet{Chernyakova2005} and \citet{Krivonos2015}, all at a flux level of $\sim$$10^{-11}$\,\flux (0.5--10 keV). Group2021 also exhibits a peculiar feature in phases 0.80--0.95. The pulse profile of Group2024, which includes the NuSTAR2024 observation and \swift\ observations at the new low state $\sim10^{-12}$ \flux, demonstrates a single-peaked shape, similar to other observations at higher fluxes. The pulse profiles are shown in Fig.~\ref{fig:period}, and the fluxes were normalized by their mean values.

\begin{figure} 
\centering
\includegraphics[width=0.9\linewidth]{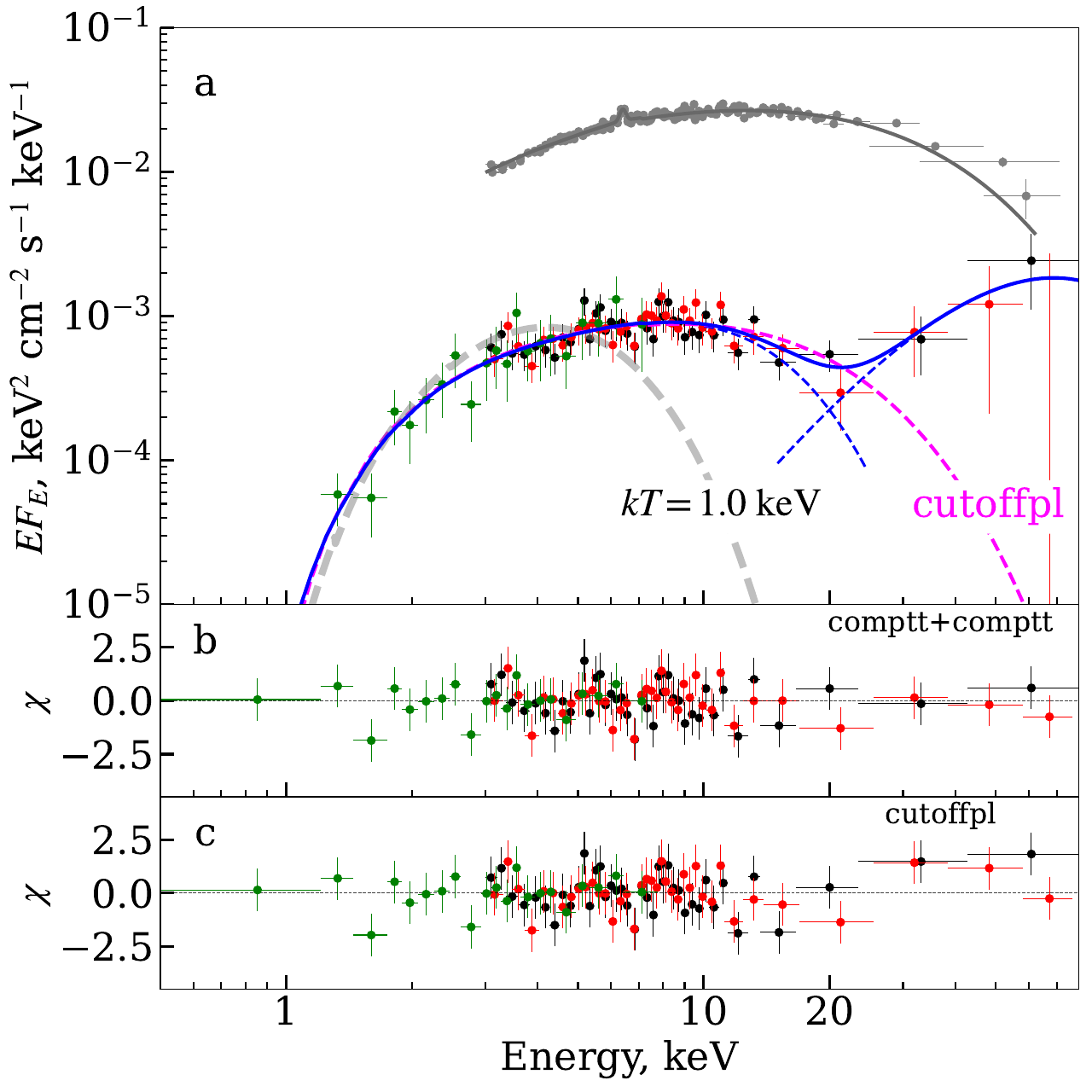}
\caption{Unfolded spectrum of \src. 
Red and black dots are for the FPMA and FPMB telescopes of the \nustar observatory (NuSTAR2024), green for the \swift/XRT telescope. 
Grey dots represent NuSTAR2014 observation   \citep{Krivonos2015}.
The blue solid line in panel (a) shows \texttt{comptt+comptt} spectral model, while the blue dashed lines represent two components separately. 
The dashed magenta line shows the best fit with the absorbed cutoff power law model.  
A blackbody model with a temperature $T=1.0~{\rm keV}$ is plotted in grey dashed line for visual comparison. 
Panels (b) and (c) show the residuals for \texttt{comptt+comptt} and \texttt{cutoffpl} continuum models, respectively.}
 \label{fig:spectrum}
\end{figure}

\subsection{Spectral analysis}
\label{sec:spectrum}

\begin{table}
\caption{Spectral parameters of \src\ from NuSTAR2024 data.}
\begin{tabular}{p{2.3cm}p{2.3cm}p{3.3cm}} 
 \hline \hline
 \small{Parameter} & \small{\texttt{tbabs$\times$cutoffpl}} & \small{\texttt{tbabs$\times$(comptt+comptt)}}\\
  \hline
 $N_{\rm H}$, $10^{22}$~cm$^{-2}$ & $2.4\pm0.6$ & $1.6^{+1.5}_{-0.7}$\\
 $\Gamma$ & $0.9\pm0.3$ & \\ 
 $E_{\rm cut}$,~keV & $7.6^{+2.7}_{-1.6}$ & \\
 $T_0$, keV & & $<0.7$ \\
 $T_{\rm e, low}$, keV & & $2.7^{+0.4}_{-0.3}$ \\
  $\tau_{\rm low}$ & & $8.3^{+1.5}_{-1.4}$ \\

  $T_{\rm e, high}$, keV & & $17^{23}_{-7}$\\
  $\tau_{\rm high}$ & & $>100$ \\

 Flux$_{2-10}$ \tablefootmark{a} & $1.9\pm0.1$ & $1.8^{+0.2}_{-0.1}$\\
 Flux$_{\rm bol}$ \tablefootmark{b} & $3.2\pm0.3$  & $5.7^{+3.8}_{-1.8}$\\
 Luminosity$_{\rm bol}$ \tablefootmark{c} & $1.1_{-0.5}^{+1.3}$ & $2.0_{-1.0}^{+2.3}$  \\
  W-stat/d.o.f. &  836/900 & 826/897  \\
 \hline
\end{tabular}
\tablefoot{\tablefoottext{a}{Flux in the 2--10 keV range in units $10^{-12}$ \flux.}
\tablefoottext{b}{Flux in the 0.5--100 keV range in units $10^{-12}$ \flux.}
\tablefoottext{c}{Luminosity in the 0.5--100 keV range in units $10^{34}$\,\lum; errors are given accounting for distance uncertainties.}
}
\label{table:spectrum}
\end{table}

The new `lowest' accretion state, characterized by luminosities ten times lower than those previously considered quiescent, is of particular interest. Figure~\ref{fig:spectrum} presents the phase-averaged energy spectrum of \src\ in this state. The NuSTAR2024 spectral data (FPMA and FPMB modules) were simultaneously fitted with data from the \swift/XRT telescope in PC mode to extend the energy range. To improve the statistical quality of the fit, \swift\ observations from 2024 July 15 to 20 (ObsIDs: 00097301021, 00097781001, 00097781003, 00097781004, 00016683005, 00097301022), which were taken close to the time of NuSTAR2024 and exhibited similar flux levels, were combined into a single spectrum.

First we tried to describe the XRP's spectrum in this state by a standard absorbed power-law with an exponential cutoff model: \texttt{tbabs $\times$ cutoffpl}. The results of the spectral fit are presented in Table~\ref{table:spectrum}. The \texttt{tbabs} component models photoabsorption using abundances from \citet{Wilms2000}. To account for the non-simultaneity of the observations and potential calibration discrepancies between FPMA, FPMB, and XRT, cross-calibration multiplicative factors were applied using the \texttt{const} model. 

The spectrum is  hard, significantly harder than a blackbody  (cf.  the blackbody spectrum of temperature $kT = 1$~keV in Fig.~\ref{fig:spectrum}). The bolometric luminosity (0.5--100~keV) was determined to be (0.6--2.4)$\times 10^{34}$\,\lum, assuming a distance range of 3.9--7.9~kpc.

However, at energies above 20~keV, the residuals of the spectrum exhibit a wavelike structure (see Fig.~\ref{fig:spectrum}c). The shape of the spectrum itself shows a deviation from a smooth form around 20~keV, which was not observed in higher luminosity states. Such behavior is typical when the spectrum has a double-humped shape, commonly seen when the XRP's luminosity falls within the $10^{34}$--$10^{36}$\,\lum\ range \citep[see, e.g., GX\,304$-$1;][]{Tsygankov2019a}. To account for this, we used a two-component Comptonization spectral model  \texttt{comptt+comptt}  \citep{Titarchuk1994}, which has been previously applied to similar XRP spectra (e.g., X\,Persei, \citealt{Doroshenko2012}; GX\,304$-$1, \citealt{Tsygankov2019a}; A\,0535$+$262, \citealt{Tsygankov2019b}; KS\,1947$+$300, \citealt{Doroshenko2020}; SRGA J124404.1$-$632232, \citealt{Doroshenko2022}; GRO\,J1008$-$57, \citealt{Lutovinov2021};  RX\,J0440.9$+$4431, \citealt{Salganik2023}).

The peaks of the \texttt{comptt+comptt} model, located at approximately 10~keV and 70~keV (plasma temperatures of  $T_{\rm e, low} = 2.7^{+0.4}_{-0.3}$~keV and $T_{\rm e, high}= 17^{+23}_{-7}$~keV, respectively), align well  with the typical positions of the soft ($\leq$10~keV) and hard ($\geq$20~keV) humps observed in other XRPs. The optical depth of the high-energy hump was fixed at $\tau = 100$ because it could not be constrained (see references above). The temperatures of the seed photons, $T_0$, for both humps were set equal to each other. The bolometric luminosity obtained using this model is slightly higher than that estimated with the \texttt{cutoffpl} continuum model: $L_{\rm bol} = (2.0_{-1.0}^{+2.3}) \times 10^{34}$\,\lum.

None of the continuum models showed a significant detection of the Fe~K$\alpha$ line, reported at higher luminosities \citep{Krivonos2015}. The 3$\sigma$ upper limit on the photon number flux for a narrow iron line ($\sigma = 0.1$~keV) at 6.4~keV is $F_{\text{iron}} = 1.8 \times 10^{-6}$ photon\,cm$^{-2}$\,s$^{-1}$, corresponding to an equivalent width of $<$0.12~keV for the double Comptonization model. No CRSFs are required for the spectrum description in either of the models. The measured neutral hydrogen column density $N_{\rm H}$ is consistent with the value observed at higher fluxes, $2.48 \times 10^{22}$~cm$^{-2}$ from \citet{Chernyakova2005}. 

\section{Discussion}

Our study of \src\ has revealed intriguing changes in its timing and spectral properties. After a long-term spin-up, the source has transitioned to an equilibrium state, marking a significant shift in its accretion behavior. This transition coincided with the emergence of a new low-luminosity accretion regime at $L_{\rm bol} \sim 10^{34}$\,\lum, an order of magnitude lower than its previously known quiescent state. The source now exhibits bimodal accretion behavior, switching between two stable luminosity states. In this newly identified low state, the spectral shape of \src has also changed, developing a two-hump structure instead of the typical cutoff power-law spectrum observed at higher luminosities. Additionally, the pulsations were detected in this new low state. Below, we discuss the implications of these findings, including estimates of the neutron star magnetic field and the nature of this previously unobserved accretion state.

\subsection{A new low-luminosity state in \src and bimodal accretion}
\label{sec:low}
One of the key findings of this study is the identification of an accretion state at $L_{\rm bol} \sim 10^{34}$\,\lum, a new for this source (see Sect.~\ref{sec:long_lc}), and an order of magnitude lower than the previously recognized quiescent luminosity of $\sim 10^{35}$\,\lum \citep{Chernyakova2005, Krivonos2015}. This newly discovered state appears to be quite stable, with the source  remaining in it for prolonged time periods. However, \src\ occasionally transitions back to the level of $L_{\rm bol} \sim 10^{35}$\,\lum, demonstrating a well-pronounced bimodal flux distribution. One of these transitions appears to exhibit the classical fast-rise exponential-decay (FRED) outburst profile characteristic of X-ray binaries, as observed at MJD~60471 (see Fig.~\ref{fig:equilibrium}e). Other transitions were characterized by a sharp rise and decline, as seen at MJD~58044 and 59280 (see Fig.~\ref{fig:equilibrium}c,d). Particularly interesting is that the transition from a luminosity state of $\sim 10^{35}$\,\lum to $\sim 10^{34}$\,\lum on MJD~58044 occurred in less than 2.3~days. Such rapid drops in luminosity are characteristic of the transition to the propeller regime in XRPs \citep[see][for  the cases of 4U~0115+63 and V~0332+53]{Tsygankov2016}. 
However, the Group2024 observations were conducted during this newly identified low-luminosity state, where the detection of pulsations provided clear evidence of ongoing accretion. This conclusion is further supported by the observed two-humped spectral shape, consistent with theoretical predictions for low-level accretion \citep{SokolovaLapa2021, Mushtukov2021}, see Sect.~\ref{sec:twohump}. Thus, these findings definitively rule out the possibility of \src\ transitioning to the propeller regime. 

This new state is highly unusual when compared to classical XRP behavior and reveals a previously unexplored phenomenon. The typical behavior of long-period BeXRPs is characterized by a stable low-luminosity state at $10^{34} - 10^{35}$\,\lum (the cold-disk accretion regime, see \citealt{Tsygankov2017}) accompanied by periodic Type I outbursts and occasional giant Type II outbursts. In contrast, the behavior observed here departs from this norm, exhibiting no stable state and instead featuring abrupt transitions between luminosity levels of $10^{34}$ and $10^{35}$\,\lum.

Given the atypical nature of these transitions, it is worth considering whether this behavior is an intrinsic peculiarity of \src\ or if it has gone unnoticed in other BeXRPs. Many persistent BeXRPs exhibit long pulse periods and low X-ray luminosities, making them difficult targets for frequent monitoring. The identification of a previously unknown accretion regime in \src\ raises the possibility that similar transitions could be occurring in other sources but have not been detected due to limited observational coverage. A systematic, high-cadence monitoring program for other persistent BeXRPs would be required to determine how common such bimodal accretion behavior is.

\subsection{Two-hump spectral shape}
\label{sec:twohump}
Previous observations of \src\ at luminosities in the range of $L_{\rm bol} \sim 10^{35}-10^{36}$\,\lum have shown that its spectrum is best described by a power-law continuum with an exponential cutoff at higher luminosities \citep{Chernyakova2005, Krivonos2015}, which is typical for accreting XRPs \citep{Coburn2002, Filippova2005}. However, the NuSTAR2024 observation, conducted at a significantly lower luminosity of $L_{\rm bol} = (2.0_{-1.0}^{+2.3}) \times 10^{34}$\,\lum, reveals a marked change in the spectral shape. Specifically, the observed spectrum is best described by a two-hump Comptonization model (see Sect.~\ref{sec:spectrum}), which includes an additional high-energy component not accounted for by the cutoff power-law model. This shift from a cutoff power-law spectrum at higher luminosities to a two-hump structure at lower luminosities is consistent with broadband observations of similar spectral transitions in other long-period XRPs, such as GX\,304$-$1 \citep{Tsygankov2019a}, A\,0535$+$262 \citep{Tsygankov2019b}, and GRO\,J1008$-$57 \citep{Lutovinov2021}. 

The two-hump spectral shape observed at low luminosities cannot be explained by the standard spectral models used for higher accretion rates. The hard X-ray component emerging at low mass accretion rates may result from cyclotron emission reprocessed by magnetic Compton scattering, while the soft component likely originates from Comptonized thermal radiation \citep{SokolovaLapa2021, Mushtukov2021}. Additionally, resonant scattering of photons by hot electrons leads to Doppler shifts in photon energies within the Doppler core of the CRSF. As a result, many photons escape the atmosphere through the wings of the line, where the cross-section is lower, forming the observed high-energy component. Consequently, the CRSF appears near the peak of the high-energy hump (as seen in A\,0535$+$262; \citealt{Tsygankov2019b}).  

Based on this, the estimated magnetic field strength is $\gtrsim 10^{13}$ G. However, this whole picture depends on the assumption that the two-hump spectral model accurately describes accretion at these low luminosities. If this model turns out to be oversimplified or if CRSF forms differently in this regime, then our magnetic field estimate could be off. Future hard X-ray observations extending beyond \nustar’s range will be needed in order to test this interpretation.

\subsection{Transition to the equilibrium state}
\label{sec:equil}

According to the current understanding of the rotational evolution of XRPs, stably accreting XRPs eventually achieve a balance between spin-up and spin-down torques, leading to what is known as the equilibrium period. As discussed in Sect.~\ref{sec:timing}, following a 20-year-long spin-up phase, \src\ has entered an ongoing 8-year phase of a stable spin period, marked by a significant change in its long-term light curve behavior. 

This transition clearly indicates that the source has reached the equilibrium state, allowing us to estimate the neutron star's magnetic field by considering the observed spin period after the transition as the equilibrium period ($P_{\rm eq}\approx 634$~s). We apply the accretion disk scenario (see, e.g., Chapter 5 in \citealt{Lipunov1992}) with fastness parameter $\omega_{\rm s} = 0.35$ \citep{GhoshLamb1979}  as follows:
\begin{equation}
\label{eq:period_disk}
P_{\rm eq} \simeq 5.7\,\Lambda^{3/2}B_{12}^{6/7}L_{37}^{-3/7}m_{\rm NS}^{-2/7}R_{6}^{15/7}~\mbox{s}. 
\end{equation}

However, it is important to note that the Ghosh \& Lamb model, while widely used, has known limitations and uncertainties. The model relies on assumptions about the disk-magnetosphere interaction that may not fully capture the complexities of real accretion processes. For example, the model assumes significant field penetration into the disk, which may be unrealistic due to high disk conductivity \citep[see][for review]{Lai2014}. It was also suggested that the accretion flow in strongly magnetized neutron stars may differ significantly from the classical Ghosh \& Lamb picture, with alternative models proposing a smaller magnetospheric radius and different torque contributions \citep{Wang1995, Bozzo2018}. Thus, while Eq.~\eqref{eq:period_disk} provides a useful framework for interpreting our results, the inferred values should be considered approximate and subject to the theoretical uncertainties.

Using the average 0.5--10~keV luminosity after the transition to equilibrium, 
$L_{\rm eq} = (4$--$17) \times 10^{34}$\,\lum (taking into account uncertainties in a distance of 
$5.4^{+2.5}_{-1.5}$~kpc), we estimate the magnetic field strength to be 
$B \approx (6$--$12) \times 10^{13}$~G. This value is in good agreement with 
the estimate derived from the power spectrum break frequency, as detailed in 
Appendix~\ref{sec:br}, and also consistent with the magnetic field strength 
inferred from the observed two-hump spectral shape, which suggests 
$B \gtrsim 10^{13}$~G (see Sect.~\ref{sec:twohump}).

\section{Conclusions}
\label{sec:sum}

We analyzed here the long-term behavior of the Be/X-ray pulsar \fullsrc, focusing on its transition to the equilibrium state. The data from \nustar, \swift, \xmm, and \chandra observatories were used to investigate changes in the XRP's timing and spectral characteristics.

We were able to establish the presence of pulsations in three observations on MJD 57511.4--57511.6, 59285.6--59309.2, and 60507.3--60520.0, demonstrating that after more than 20 years of continuous spin-up, \src has reached a stable spin period $P_{\rm eq}$ of approximately 634~s, indicating a shift to an equilibrium state. All three observations demonstrated single-peaked pulse profiles. This transition coincided with the emergence of a previously unknown low-luminosity accretion regime at $L_{\rm bol} \sim 10^{34}$\,\lum, an order of magnitude lower than its previously known quiescent state. Additionally, we find tentative evidence for an orbital period, $P_{\rm orb} \approx 90$~days, based on three consecutive outbursts around MJD 60470, 60560, and 60650.

The source exhibited behavior atypical for BeXRPs, occasionally transitioning between these two stable luminosity levels of approximately $10^{34}$\,\lum and $10^{35}$\,\lum, remaining in both states for extended periods -- well-pronounced bimodal behavior. 
The transition from $10^{35}$\,\lum to $10^{34}$\,\lum occurred in less than 2.3~days (the interval between consecutive observations, though the transition itself may have taken an even shorter time). 
This behavior could indicate a transition to the propeller regime. 
However, timing analysis of this state confirmed the presence of pulsations, while spectral analysis revealed a two-hump Comptonization shape, in contrast to the cutoff power-law spectrum observed at higher luminosities. 
The detection of pulsations and the two-hump spectral shape excludes the possibility of a propeller regime in this newly identified state. 
The mechanism driving the transition to this previously unknown accretion regime remains unclear. However, the identification of a bimodal accretion behavior in \src\ suggests that similar transitions may occur in other sources but have remained undetected due to limited observational coverage.

Despite the absence of CRSF detection in the spectrum, we estimated the magnetic field strength using the equilibrium spin period value and using the perturbation propagation model. 
In these two estimations, the magnetic field was of the order of $\simeq10^{13}-10^{14}$~G, which agrees well with the observed spectral shape in the low state. Note that for such a strong magnetic field, CRSF should indeed not be observable in the \nustar energy range.

However, it is important to note that these estimates are strongly model-dependent and subject to significant uncertainties. 
Future observations, particularly in a broader X-ray energy range, will be crucial for refining these estimates and further understanding the accretion mechanisms in this system.

\begin{acknowledgements}
AS acknowledges support from the EDUFI Fellowship and Jenny and Antti Wihuri Foundation. DM acknowledge support by the state of Baden-W\"urttemberg through~bwHPC.
We are grateful to the \swift team for approving and rapid scheduling of the monitoring campaign. We are grateful to the \nustar team for approving and rapid scheduling of the observation. This work made use of data supplied by the UK Swift Science Data Centre at the University of Leicester and data obtained with NuSTAR mission, a project led by Caltech, funded by NASA, and managed by JPL. 
This research also has made use of the \nustar Data Analysis Software (NUSTARDAS) jointly developed by the ASI Science Data Centre (ASDC, Italy) and Caltech. 
This research has made use of data and software provided by the High Energy Astrophysics Science Archive Research Centre (HEASARC), which is a service of the Astrophysics Science Division at NASA/GSFC and the High Energy Astrophysics Division of the Smithsonian Astrophysical Observatory. 
This research is based on observations obtained with \xmm, an ESA science mission with instruments and contributions directly funded by ESA Member States and NASA. 
The scientific results reported in this article are based in part of data obtained from the \chandra Data Archive.

\end{acknowledgements} 

\bibliography{allbib.bib}{}
\bibliographystyle{aa}

\begin{appendix}

\section{Disk truncation}
\label{sec:br}

The light curves of XRPs exhibit both periodic and aperiodic variability. A widely accepted model for describing the aperiodic variability is the `perturbation propagation' model \citep{Lyubarskii1997}. In this model, variations in X-ray brightness are attributed to fluctuations in the mass accretion rate on Keplerian timescales, driven by the multiplicative superposition of stochastic perturbations within the accretion flow. These perturbations originate from viscous stresses at different radii in the disk. When integrated, the resulting variability produces a characteristic red noise with a power-law power spectrum. 

The power spectrum $P(f)$ breaks at the break frequency $f_{\rm br}$, which corresponds to the Keplerian frequency $\nu_{\rm K}$ at the inner disk radius. 
As a result, below $f_{\rm br}$, $P(f) \propto f^{-1}$ \citep{Lyubarskii1997}, and at higher Fourier frequencies, the spectrum is $P(f) \propto f^{-2}$.
Measurement of the break frequency allows the estimation of the inner disk radius $R_{\rm m}$ and thus the magnetic field \citep{Revnivtsev2009}: 
\begin{equation}\label{eq:fbrnuK}
2 \pi f_{\rm br} = 2 \pi \nu_{\rm K} =  \left( GM_{\rm NS} \right)^{1/2} R_{\rm m}^{-3/2} . 
\end{equation} 
We can relate the magnetospheric radius $R_{\rm m}$ to the luminosity and the magnetic field strength as \citep[e.g.,][]{MushtukovTsygankov2024}: 
\begin{equation} \label{eq:Rm}
R_{\rm m} \simeq 2.4 \times 10^8 \Lambda B_{12}^{4/7} {L}_{37}^{-2/7} m_{\rm NS}^{1/7} R_6^{10/7}~\text{cm},
\end{equation} 
where $m_{\rm NS}=M_{\rm NS}/M_\odot$, 
$L_{37}$ is the luminosity in units of $10^{37}$\,\lum, $R_{6}$ is the neutron star radius in $10^6$~cm, and $B_{12}$ is the neutron star magnetic field in $10^{12}$~G. 
The coefficient $\Lambda$ relates the Alfv\'en radius to the magnetospheric radius $R_{\rm m}=\Lambda R_{\rm A}$. 
For accretion from a gas pressure-dominated disk, this coefficient is typically taken to be $\Lambda=0.5$ \citep{GhoshLamb1979}.

Based on the study of \src\ using NuSTAR2014 data, \citet{Krivonos2015} measured the break frequency $f_{\rm br} = 0.0066$~Hz in the power spectrum at a bolometric flux of $1.2 \times 10^{-10}$\,\flux (calculated based on spectral parameters from \citealt{Krivonos2015}). 
From Eqs.~\eqref{eq:fbrnuK} and \eqref{eq:Rm}, we estimate the magnetic field to be $B \approx$(5--11)$\times 10^{13}$~G assuming canonical neutron star mass $M = 1.4 M_{\odot}$ and radius $R = 12$~km. 

The identification of the break frequency $f_{\rm br}$ with the Keplerian frequency at the magnetospheric radius is not firmly established and remains a subject of debate \cite[see, e.g.,][]{Monkkonen2019, Monkkonen2022}. While this assumption is commonly used, the exact origin of the broad-band noise in power spectra is not fully understood, and alternative interpretations exist. Additionally, the relation between $f_{\rm br}$ and the inner disk radius may be influenced by factors beyond magnetic truncation, such as local disk instabilities or other accretion flow properties. In light of these uncertainties, estimates of the neutron star's magnetic field based on $f_{\rm br}$ should be treated with caution, as they may be subject to systematic uncertainties arising from these alternative influences.

\section{Observation log}
\xmm and \swift/XRT observations used in this work can be found in Tables~\ref{table:xmm_obs} and \ref{table:swift_obs}.
\begin{table}[!ht]
\caption{{\xmm observations used in this work.}}
\centering
\label{table:xmm_obs}
\begin{tabular}{cc}
\hline\hline
ObsID & MJD \\
& (d)\\
\hline
0092820101 & 51921.7 \\
0092820201 & 52101.3 \\
0092820301 & 52467.2 \\
0092820801 & 52837.5 \\
0092821201 & 52668.3 \\
0201920101 & 53028.8 \\
0201920201 & 53045.4 \\
0201920301 & 53055.8 \\
0201920401 & 53053.0 \\
0201920501 & 53051.4 \\
0504550501 & 54289.6 \\
0504550601 & 54298.2 \\
0504550701 & 54329.4 \\
0653640401 & 55567.8 \\
0653640501 & 55594.9 \\
0653640601 & 55624.3 \\
0881420201 & 59226.0 \\
0881420301 & 59222.0 \\
0881420401 & 59230.0 \\
0915391301 & 60008.4 \\
\hline
\end{tabular}
\end{table}

\begin{table*}[!ht]
\centering
\caption{\swift/XRT observations used in this work.}
\label{table:swift_obs}
\begin{tabular}{cc cc cc cc}
\hline\hline
ObsID & MJD & ObsID & MJD & ObsID & MJD & ObsID & MJD \\
& (d) && (d) && (d) && (d)\\
\hline
00030966002 & 54290.6 & 00030966072 & 58018.8 & 00095944020 & 59305.1 & 00097301018 & 60503.3 \\
00030966005 & 55414.3 & 00030966073 & 58024.0 & 00095944021 & 59307.1 & 00097301019 & 60504.4 \\
00030966007 & 55416.0 & 00030966075 & 58034.9 & 00095944022 & 59309.2 & 00097301021 & 60506.3 \\
00030966008 & 55417.4 & 00030966076 & 58037.9 & 00030966119 & 59317.8 & 00097781001 & 60507.3 \\
00030966009 & 55420.0 & 00030966077 & 58039.6 & 00030966120 & 59319.7 & 00097781004 & 60509.2 \\
00030966010 & 55423.3 & 00030966078 & 58040.7 & 00030966121 & 59321.8 & 00016683005 & 60509.5 \\
00030966011 & 55520.2 & 00030966079 & 58042.0 & 00030966122 & 59323.8 & 00097301022 & 60511.5 \\
00030966012 & 55525.6 & 00030966080 & 58043.9 & 00030966123 & 59324.7 & 00016683006 & 60512.4 \\
00030966013 & 55530.5 & 00030966081 & 58046.2 & 00030966124 & 59325.7 & 00097301023 & 60513.5 \\
00030966014 & 55544.6 & 00030966082 & 58049.3 & 00030966126 & 59327.8 & 00097301024 & 60515.3 \\
00030966015 & 55580.7 & 00030966083 & 58052.2 & 00030966127 & 59329.8 & 00097781005 & 60518.1 \\
00030966016 & 55581.1 & 00030966084 & 58053.9 & 00030966128 & 59330.7 & 00016683008 & 60518.1 \\
00030966017 & 55583.7 & 00030966085 & 58056.0 & 00030966129 & 59331.5 & 00097781006 & 60519.9 \\
00030966018 & 56767.0 & 00030966086 & 58057.2 & 00095731001 & 59332.4 & 00016683009 & 60521.0 \\
00080099001 & 56767.5 & 00030966087 & 58059.0 & 00030966130 & 59333.5 & 00097301025 & 60521.2 \\
00030966019 & 56774.8 & 00030966088 & 58072.6 & 00095731002 & 59334.6 & 00097301026 & 60523.1 \\
00030966020 & 56781.9 & 00030966089 & 58075.7 & 00095731003 & 59336.4 & 00030966134 & 60524.1 \\
00030966021 & 56783.9 & 00030966092 & 58096.0 & 00095731004 & 59338.6 & 00016683010 & 60524.3 \\
00030966022 & 56784.1 & 00030966093 & 58099.0 & 00095731005 & 59340.7 & 00097301027 & 60525.0 \\
00030966023 & 56786.1 & 00030966095 & 58108.1 & 00095731006 & 59342.4 & 00030966136 & 60526.0 \\
00030966024 & 56794.6 & 00030966096 & 58110.3 & 00095731007 & 59344.6 & 00097301028 & 60527.9 \\
00030966025 & 56796.9 & 00030966097 & 58119.4 & 00095731008 & 59346.3 & 00097301029 & 60529.0 \\
00030966037 & 56823.0 & 00030966100 & 59233.9 & 00095731009 & 59348.2 & 00030966138 & 60531.2 \\
00030966040 & 56825.5 & 00030966101 & 59237.6 & 00095731010 & 59350.1 & 00030966139 & 60532.2 \\
00080748001 & 56832.1 & 00030966103 & 59239.8 & 00095731011 & 59352.4 & 00097301030 & 60544.5 \\
00030966043 & 56833.2 & 00030966105 & 59245.7 & 00095731012 & 59354.1 & 00097301031 & 60547.4 \\
00030966044 & 56834.1 & 00030966106 & 59248.6 & 00095731013 & 59356.5 & 00097301032 & 60550.4 \\
00030966047 & 56835.2 & 00030966107 & 59252.6 & 00095731014 & 59358.0 & 00097301033 & 60553.2 \\
00030966048 & 56840.4 & 00030966108 & 59256.5 & 00030966131 & 60010.4 & 00097301034 & 60556.3 \\
00030966049 & 56844.7 & 00095944001 & 59267.9 & 00030966132 & 60017.7 & 00097301035 & 60559.1 \\
00030966050 & 56845.1 & 00030966111 & 59268.6 & 00030966133 & 60024.4 & 00097301036 & 60562.1 \\
00030966051 & 56846.5 & 00095944003 & 59271.4 & 00097301002 & 60462.3 & 00016683012 & 60595.6 \\
00030966052 & 57838.7 & 00095944004 & 59273.5 & 00097301003 & 60467.6 & 00016683013 & 60598.5 \\
00030966053 & 57848.1 & 00095944005 & 59275.8 & 00097301004 & 60471.2 & 00016683014 & 60601.2 \\
00030966055 & 57850.1 & 00095944006 & 59277.6 & 00097301005 & 60472.1 & 00016683015 & 60608.8 \\
00030966056 & 57855.9 & 00030966115 & 59278.4 & 00097301006 & 60473.1 & 00016683016 & 60613.0 \\
00030966057 & 57858.4 & 00095944007 & 59279.8 & 00097301007 & 60474.2 & 00080099002 & 60615.3 \\
00030966058 & 57868.6 & 00095944008 & 59281.5 & 00097301008 & 60475.0 & 00016683017 & 60618.1 \\
00030966059 & 57987.8 & 00030966117 & 59282.8 & 00097301009 & 60476.5 & 00016683018 & 60623.1 \\
00030966060 & 57990.0 & 00095944009 & 59283.6 & 00097301010 & 60479.0 & 00016683019 & 60628.1 \\
00030966061 & 57993.1 & 00030966118 & 59284.7 & 00097301011 & 60482.3 & 00016683020 & 60633.2 \\
00030966062 & 57998.8 & 00095944010 & 59285.6 & 00097301012 & 60484.0 & 00016683022 & 60648.9 \\
00030966063 & 58000.2 & 00095944011 & 59287.7 & 00097301013 & 60493.7 & 00016683023 & 60653.9 \\
00030966064 & 58001.2 & 00095944012 & 59289.2 & 00097301014 & 60495.6 & 00016683024 & 60659.1 \\
00030966065 & 58002.4 & 00095944013 & 59291.5 & 00016683001 & 60497.4 & 00016683025 & 60664.5 \\
00030966067 & 58004.0 & 00095944014 & 59292.6 & 00097301015 & 60499.4 & 00016683026 & 60668.3 \\
00030966068 & 58005.2 & 00095944016 & 59297.4 & 00016683002 & 60500.2 & 00016683027 & 60673.6 \\
00030966069 & 58006.7 & 00095944017 & 59299.7 & 00097301016 & 60501.5 & 00016683028 & 60678.9 \\
00030966070 & 58008.0 & 00095944018 & 59301.3 & 00097301017 & 60502.3 & 00016683029 & 60683.1 \\
00030966071 & 58013.7 & 00095944019 & 59303.1 & 00016683003 & 60503.2 & & \\
\hline
\end{tabular}
\end{table*}
\end{appendix}

\end{document}